\documentclass[%
superscriptaddress,
preprintnumbers,
nofootinbib,
 amsmath,amssymb,
 aps, 
twocolumn,
prl,
]{revtex4}
\usepackage{graphicx}
\usepackage{amssymb}
\usepackage{amsbsy}
\usepackage{amsmath}
\usepackage{chngcntr}
\usepackage[export]{adjustbox}
\usepackage{color}

\newcommand {\fabsq}[1] {\left| #1 \right|^2}

\def\ra{\rangle}
\def\la{\langle}

\newcommand{\beqa}{\begin{eqnarray}}
\newcommand{\eeqa}{\end{eqnarray}}
\newcommand{\beq}{\begin{equation}}
\newcommand{\eeq}{\end{equation}}

\newcommand{\cR}{{\cal{R}}}
\newcommand{\cT}{{\cal{T}}}
\newcommand{\cE}{{\cal{E}}}
\newcommand{\cD}{{\cal{D}}}

\begin{document}
\title{Maxwell demon no-go theorem for multichannel scattering}
\author{Andreas Ruschhaupt}
\affiliation{School of Physics, University College Cork, Cork, T12 K8AF, Ireland}
\author{Juan Gonzalo Muga}
\affiliation{
IUDEA, Universidad de La Laguna, Apdo 456, La Laguna 38205, Tenerife, Spain
}
\affiliation{EHU Quantum Center, University of the Basque Country UPV/EHU, 48940 Leioa, Spain}
\begin{abstract}
Recent implementations of Maxwell demons and other devices exhibiting asymmetric response
to particles incident from 
opposite directions have raised conceptual and practical interest. According to quantum-scattering-theory selection rules,  demons
necessitate, among other conditions,  non-local and non-Hermitian potentials. These potentials
arise  as effective interactions for a single asymptotic channel  within a multichannel (large, Hermitian) system.  
However we find that, regardless of the complexity of the large Hamiltonian, the demon behavior cannot be approached in this way.
We discuss the consequences and possible resolutions of the ensuing paradox. 
We also introduce and examine a parameter to quantify the distance to demon behavior and propose a way to construct
devices as close to a demonic behaviour as possible in a quantum-optical setup.
\end{abstract}
\maketitle
%
%

%
{\bf Introduction.}
Maxwell's demon has been a catalyst to sharpen our understanding of the second law of thermodynamics and the interplay among fundamental concepts such as entropy, information, measurement, and work.  A rich history surrounds this topic, which has been thoroughly explored in various reviews and collections of articles.  Eminent physicists, including Maxwell himself, Smoluchowski, Szilard, Feynman, Landauer, and numerous others, have made substantial contributions \cite{Maxwell1990,Maxwell2002,Rex2017}. 
Prevailing tendencies in these studies were to either theoretically discuss the failure of certain demons, or to argue why the second law remains intact when they function properly. However, more recently, there has been a growing interest in the practical applications of Maxwell's demon \cite{Parrondo2021}, including microscopic devices \cite{Koski2015};  biological motors \cite{Leigh2007}, or the use of information as a fuel \cite{Mayusama2018}. Further examples of this trend are the atom diode \cite{Ruschhaupt2004a,Ruschhaupt2006a,Ruschhaupt2006b} and one-way barriers implemented in cold atom laboratories for trapping and cooling \cite{Raizen2005,Raizen2009}.  All these devices are based on the principle of asymmetric response, a characteristic feature of valves, diodes or rectifiers,
which play a pivotal role in nature and technology, including optics \cite{Sheikh2023}. It is thus
reasonable to anticipate that demons and asymmetric-response devices will assume a significant role in quantum-based technologies, integrated in atomotronic circuits
for metrology, in sensing and in quantum information processing applications, and also to implement
cooling techniques, isolators, and particle or energy harvesting. 
Their study is thus of both fundamental and practical interest.
 
Different variants of the demon have been put forward, even in Maxwell's time, 
so it is important to clearly set the specific scope of the present work.  Our primary focus will be on ``pressure demons'', and more  specifically 
on automated gate devices for one-way passage of microscopic quantum particles.
In one dimension this demon type is defined by extreme asymmetries in the transmission and reflection amplitudes for incidence from left or right. We will consider mostly a left-incidence-transmitted demon which is then defined by
\begin{equation}
|T|=|\widetilde{R}|=1,\; |\widetilde{T}|=|R|=0,
\label{MDconditions}
\end{equation}
where $T$ and $R$ are transmission and reflection amplitudes for left incidence and $\widetilde{T}$ and $\widetilde{R}$ for right incidence. 
(By interchanging the tilde and not-tilded symbols one would get a right-incidence-transmitted demon.)
For practical use these relations should typically hold for some incident-energy window, but a first step is to examine their 
possible realization for a single particular energy. One more caveat is that we assume throughout the manuscript, except when stated otherwise,  
that the dynamics of the particle is described by a Schr\"odinger equation, possibly of a very large system, although we shall focus on a subsystem (a particle in a given channel) to define the demon according to the above conditions (\ref{MDconditions}). These conditions imply that the wave remains quantum-mechanically
coherent, in principle a desirable goal in circuits designed for quantum-technology applications. We shall later discuss relaxations of these constraints.              

Our analysis will lead us to establish a  no-go theorem, which rules out the possibility of the subsystem to even approximate a behavior reminiscent of a demon; we study this  in a quantitative way.
We shall finally discuss the consequences and potential resolutions that arise from this paradoxical scenario.

%
{\bf Asymmetric scattering devices.}
Scattering theory in one dimension (1D) \cite{Muga2004,Mosta2014},
provides symmetry-based selection rules \cite{Ruschhaupt2017,Simon2018,Simon2019,Ruschhaupt2021} that tell us if
the conditions in (\ref{MDconditions}), 
or in fact any other form of asymmetric response are at all possible for a given 1D, single-channel Hamiltonian with kinetic and potential terms, 
$H = \frac{P^2}{2m} + V(X)$, for a particle of mass $m$ moving along the $X$ axis. Nonlinear effects may violate these rules \cite{Lin2011,Peng2014,Mosta2019,Mekawy2021}, but here linear dynamics is assumed.
 
A well-known fact is that a Hermitian (1D, single-channel) Hamiltonian is not compatible with the demon conditions in (\ref{MDconditions}). 
Hermiticity is one among eight symmetries identified in refs. \cite{Ruschhaupt2017,Simon2018,Simon2019,Ruschhaupt2021} as the invariance of $H$ with respect to transformations represented by superoperators  for complex conjugation, inversion and transposition \cite{Simon2019}. These are unitary and antiunitary symmetries that slightly generalize Wigner's definition of a symmetry \cite{Wigner1959}, in the sense that they maintain the transition probabilities,  i.e., the overlap, among density operators, rather than among pure states as in Wigner's standard definition \cite{Simon2018,Ruschhaupt2021}.   
In fact none of the seven non-trivial symmetries allows for a demon \cite{Ruschhaupt2017},  this includes parity, parity-time (PT), time-reversal invariance, 
time-reversal pseudohermiticy, parity pseudohermiticity, and PT-pseudohermiticity, see Table \ref{ta}. We are left with the trivial symmetry as the only possible niche to find a demon.      
\begin{table}
\begin{center}
\begin{tabular}{ll}
{Symmetry\; name}&Matrix\; elements\\
\hline
{\rm Trivial}&
$\langle x|V|y\rangle=\langle x|V|y\rangle$ 
\\
{\rm Hermiticity}&
$\langle x|V|y\rangle=\langle y|V|x\rangle^*$ 
\\
{\rm Parity}&
$\langle x|V|y\rangle=\langle -x|V|\!-\!y\rangle$ 
\\
{\rm Parity\; pseudohermiticity}&
$\langle x|V|y\rangle=\langle -y|V|\!-\!x\rangle^*$ 
\\
 {\rm Time\; reversal\; invariance}&
 $\langle x|V|y\rangle=\langle x|V|y\rangle^*$
\\
{\rm Time\; reversal\; pseudohermiticity}&
$\langle x|V|y\rangle=\langle y|V|x\rangle$
\\
{\rm PT\; symmetry}&
$\langle x|V|y\rangle=\langle -x|V|\!-\!y\rangle^*$
\\
{\rm PT\; pseudohermiticity}&
$\langle x|V|y\rangle=\langle -y|V|\!-\!x\rangle$ 
\end{tabular}
\end{center}
\vspace*{-0.5cm}
\caption{Symmetries of potential matrix elements in coordinate representation\label{ta}}
\vspace*{-0.5cm}
\end{table}
Note in particular that parity, time-reversal pseudohermiticity, or parity-time symmetry do not allow for asymmetric transmission so they do not provide demons \cite{Ruschhaupt2017}. Time-reversal pseudohermiticity  is satisfied by any local potential in coordinate space 
(whose non-vanishing elements are of the diagonal form $\la x|V|x\ra$), so the restrictions are quite severe.  
Non-local, non-Hermitian Hamiltonians leading to demons as defined by (\ref{MDconditions}) have been described and found by simple inversion methods (see the Supplementary Material in ref. \cite{Ruschhaupt2017}). A compelling question then is if these demon-consistent formal potentials could be realized physically.    
 
Non-local and non-Hermitian Hamiltonians arise naturally when applying Feshbach's projection technique \cite{Feshbach1958,Feshbach1962} to find an effective Hamiltonian for a given channel from a larger, multichannel Hamiltonian $H_L$, hereafter the ``large Hamiltonian''. 
A simple example is an atom with internal states affected in a localized region by a laser. The different channels are 
then represented by the internal states and it is possible to find effective non-local and non-Hermitian Hamiltonians for the subspace spanned asymptotically (i.e., before and after the interaction) by one of them, for example the ground state \cite{Ruschhaupt2004b,Ruschhaupt2020}. Whereas some asymmetric devices can indeed be found via this quantum-optical
realization, it was shown that a demon, with the amplitudes for the selected channel satisfying (\ref{MDconditions}),
cannot be found this way \cite{Ruschhaupt2020}. There are of course other physical systems that may be described, at least in principle,  in a multichannel
scenario. For example, the moving particle may be coupled to other (environment) particles in a localized interaction region and the channels 
would be associated with their excitations. Frictional effects, in particular, are often modeled by these interactions. 
It is thus of interest to determine in a generic way if a demon behavior can be found for one channel if the underlying dynamics is driven by an arbitrary large Hermitian Hamiltonian.

%
{\bf No-go theorem.}
Let us consider a quantum particle moving on the $x$ axis. 
This particle has internal levels $|a\rangle$ 
that are only coupled to each other in a localized region of space by a potential $V_L$, a matrix with respect to internal levels. 
An alternative is that the particle
does not have a structure but interacts with a quantum system with discrete levels $|j\rangle$ in a localized region.   
In any case we write the large Hamiltonian as  
$H_L = \frac{P^2}{2m} + V_L +H_{\rm int}$, where $V_L$  decays fast enough in the coordinate representation of the moving particle to apply the completeness relations of scattering theory \cite{Muga2004} and the internal Hamiltonian is
$H_{\rm int}=\sum |j\rangle \epsilon_j \langle j|$.  Each asymptotic channel is characterized by 
an internal state.
For a given {\em open} channel, where $E> \epsilon_j$, there are two incoming and two outgoing subchannels corresponding
to rightwards and leftwards motion with momentum $\hbar k_j$ resp. $-\hbar k_j$.  The total energy in the channel is distributed
among kinetic and internal energies,
$
E=k_j^2 \hbar^2/(2m) +\epsilon_j.
$
A channel is closed if $E< \epsilon_j$, which implies that it does not carry any asymptotic flux. In a time-dependent process closed channels may provide a transient amplitude to the wavefunction. In a stationary scattering wave, closed channels cannot contribute to the   
asymptotic incoming or outgoing waves.   
A wave incident from the left in state $|i\rangle$ takes the asymptotic form (still a vector in internal state space)
$
|\pmb{\psi}_i (x)\ra \underset{x\sim -\infty}{\sim}   \frac{e^{i k_i x}}{\sqrt{k_i}} |i\rangle 
\!+\!  
\sum_{j\, {\rm open}} R_{j,i} \frac{e^{-i k_j x}}{\sqrt{k_j}} |j\rangle
, 
|\pmb{\psi}_i (x)\ra \underset{x\sim \infty}{\sim}
\sum_{j\, {\rm open}} T_{j,i} \frac{e^{i k_j x}}{\sqrt{k_j}} |j\rangle
$
where $k_i,k_j>0$. A wave incident from the right in state $|i\rangle$ takes the asymptotic form
$
|\tilde{\pmb{\psi}}_i (x)\ra \underset{x\sim \infty}{\sim}   \frac{e^{-i k_i x}}{\sqrt{k_i}} |i\rangle
\!+\!
\sum_{j\, {\rm open}} \tilde R_{j,i} \frac{e^{i k_j x}}{\sqrt{k_j}} |j\rangle
,
|\tilde{\pmb{\psi}}_i (x)\ra \underset{x\sim -\infty}{\sim}
\sum_{j\, {\rm open}} \tilde T_{j,i} \frac{e^{-i k_j x}}{\sqrt{k_j}} |j\rangle
$
where again $k_i,k_j>0$.
The plane waves are normalized by factors $1/\sqrt{k_j}$ (outgoing) and $1/\sqrt{k_i}$ (incoming) so that the modulus squared scattering amplitudes represent probabilities
for the corresponding outgoing subchannel.   
 
The ${\sf S}$-matrix for $N$ open channels corresponding to $N$ internal states and the two directions is a $2N \times 2N$ matrix: 
\beq
{\sf S}=\left(\begin{array}{cc}
{\sf T} &\widetilde {\sf R}
\\
{\sf R} &\widetilde {\sf T}
\end{array}\right) .
\eeq
${\sf T}$ and ${\sf R}$ are $N\times N$ matrices of transmission and reflection amplitudes 
for left incidence, whereas ${\sf \widetilde{T}}$ and ${\sf \widetilde{R}}$ are the corresponding matrices 
for incidence from the right.

The ${\sf S}$ matrix is unitary for Hermitian Hamiltonians, which reflects the conservation of total flux between incoming 
and outgoing 
waves. The unitarity relations ${\sf S}^\dagger {\sf S}={\sf S} {\sf S}^\dagger={\sf 1}_{2N}$ give in particular
\begin{eqnarray}
{\sf T}^\dagger {\sf T} + {\sf R}^\dagger {\sf R} &=& {\sf 1}_N,\;\;
\widetilde {\sf T}^\dagger \widetilde {\sf T} + \widetilde {\sf R}^\dagger \widetilde {\sf R} = {\sf 1}_N,
\\
{\sf T} {\sf T}^\dagger + \widetilde{\sf R} \widetilde {\sf R}^\dagger &=& {\sf 1}_N,\;\; 
\widetilde {\sf T} \widetilde {\sf T}^\dagger + {\sf R} {\sf R}^\dagger = {\sf 1}_N.
\end{eqnarray}
For $i=j$ we get 
\begin{eqnarray}
\sum_k \left( |T_{k,i}|^2 + |R_{k,i}|^2 \right) =1,
\sum_k \left( |\widetilde T_{k,i}|^2 + |\widetilde R_{k,i}|^2 \right) = 1,
\label{eqa}\\
\sum_k \left( |T_{i,k}|^2 + |\widetilde R_{i,k}|^2 \right) = 1
,
\sum_k \left( |\widetilde T_{i,k}|^2 + |R_{i,k}|^2 \right) = 1,\label{eqb}
\end{eqnarray}
Eq. \eqref{eqa} (left) can be rewritten as $|T_{i,i}|^2 + |R_{i,i}|^2  + \sum_{k\neq i} \left( |T_{k,i}|^2 + |R_{k,i}|^2 \right) =1$ and so it follows
that $|T_{i,i}|^2 + |R_{i,i}|^2 \le 1$. In summary, the relations \eqref{eqa}-\eqref{eqb} lead to the following restrictions for the  amplitudes within the $i$-th channel:
\beqa
|T_{i,i}|^2 + |R_{i,i}|^2 &\le& 1,\;\;
|\widetilde T_{i,i}|^2 + |\widetilde R_{i,i}|^2 \le  1,
\label{a2}\\
|T_{i,i}|^2 + |\widetilde R_{i,i}|^2 &\le& 1 ,\;\;
|\widetilde T_{i,i}|^2 + |R_{i,i}|^2 \le  1 .
\label{a4}
\eeqa
While the first two inequalities  in (\ref{a2})  are rather obvious because of  probability conservation for an incoming wave, the last two equations in 
(\ref{a4}) are less so as they connect amplitudes for left and right incidence, but they also result from flux conservation due to ${\sf S}{\sf S}^\dagger=1$, and
set physical  limits to the possible asymmetric devices that can be constructed for a given channel.
The extreme constraints following by requiring a probability to be one based on these inequalities (\ref{a2}-\ref{a4})
are summarized in Table \ref{table2} for the four different cases.

\begin{table}
\begin{center}
\begin{eqnarray*}
\begin{array}{c|c|c|c}
|T_{i,i}|^2 & |R_{i,i}|^2 & |\widetilde{T}_{i,i}|^2 & |\widetilde{R}_{i,i}|^2\\ \hline
{\pmb 1} & 0 & \le 1 & 0 \\
0 & {\pmb 1} & 0 & \le 1 \\
\le 1 & 0 & {\pmb 1} & 0 \\
0 & \le 1 & 0 & {\pmb 1}
\end{array}
\end{eqnarray*}
\end{center}
\vspace*{-0.5cm}
\caption{\label{table2}Constraints following from requiring a single probability to be one (bold) based on the inequalities (\ref{a2}-\ref{a4}).}
\vspace*{-0.5cm}
\end{table}

We are considering now asymmetric devices which are based on a selected internal subspace $i_0$ and let
$T=T_{i_0,i_0}, R=R_{i_0,i_0}, \widetilde T = \widetilde{T}_{i_0,i_0}, \widetilde R = \widetilde{R}_{i_0,i_0}$.
Devices
with extreme asymmetric scattering response
may be classified by a two sided letter code,  with letters on the left of a dash describing the effect of left incidence, and letters on the right
the effect of incidence from the right. The letters used are ${\cal T, R, A}$,  denoting full transmission, full reflection, and full "absorption" respectively \cite{Ruschhaupt2017}.
Considering the multichannel large system perspective, absorption is attributed here to the transition of a particle from the incident channel $i_0$ to a different channel $i \neq i_0$.
Note that ``full" transmission and reflection are characterized by  
a scattering amplitude with modulus one. As the large Hamiltonian is assumed Hermitian, only combinations that 
conserve or ``absorb''
the incoming flux are in principle allowed, such as ${\cal T/R}$, ${\cal T/A}$, or ${\cal R/A}$, which are analogous to different forms of rectifiers in electronic circuits.
(Other combinations were considered in \cite{Ruschhaupt2017} if the non-Hermitian Hamiltonian is
not derived from a large Hermitian Hamiltonian.)

In this notation the demon is a ${\cal T/R}$ device, i.e.
we require \eqref{MDconditions}.
To quantify how close we can get to a Maxwell demon, we define the parameter
\begin{eqnarray}
\cD = \frac{1}{2} \left[|T|^2 + \left(1-|R|^2\right) 
+ |\widetilde R|^2 + \left(1-|\widetilde T|^2\right)\right] - 1.
\end{eqnarray}
As all probabilities $|T|^2, |R|^2, |\widetilde T|^2, |\widetilde R|^2$ are $\ge 0$
and also $\le 1$, we get immediately the following bounds for this parameter: $-1 \le \cD \le 1$. It also follows that
$\cD=+1$ corresponds to the fulfillment of the requirements \eqref{MDconditions} for a left-incidence-transmitting demon ${\cal T/R}$. In a similar way,
it follows that $\cD=-1$ correspond to a right-incidence-transmitting demon ${\cal R/T}$.

However, the inequality \eqref{a4} (right) implies a stronger upper bound to this parameter, namely
$\cD \le \frac{3}{2} - 1 = \frac{1}{2}$, and the inequality \eqref{a4} (left) implies a stronger lower
bound, namely $\cD \ge \frac{1}{2} - 1 = -\frac{1}{2}$. In summary, $-\frac{1}{2} \le \cD \le \frac{1}{2}$.
This means that the inequalities \eqref{a4} forbid the one-way barrier ${\cal T/R}$, Eq. \eqref{MDconditions}, as well as ${\cal R/T}$.
This no-go theorem is a remarkably general result because it is independent of the dimension of the Hilbert space! Even for an arbitrary large Hilbert space resp. Hamiltonian, one cannot even get close to a Maxwell demon!

This can be also seen directly by the fact that the one-way barrier ${\cal T/R}$ would need
\begin{eqnarray}
|T|^2 + |\widetilde R|^2 = 2 &\not\le& 1 ,\;\;
\end{eqnarray}
a contradiction with the first equation in (\ref{a4}).

It is important to underline that the bounds $-\frac{1}{2} \le \cD \le \frac{1}{2}$ even forbid devices ``close" to a Maxwell demon.
For example, $\fabsq{T} > 1/2$ together with $\fabsq{\widetilde{R}} > 1/2$ would result in $\cD > \frac{1}{2}$
which is forbidden.

Let us examine the effect of the symmetries listed in Table \ref{ta}. In detail, we examine the symmetries of the effective potential of channel $i_0$ obtained by applying Feshbach's projection technique \cite{Feshbach1958,Feshbach1962} with
$P=|i_0\ra\la i_0|$ being the projector on channel $i_0$ and $Q=1-P$.
The effective potential for channel $i_0$ then becomes
$
V=P V_L P+P V_L Q(E+i0-Q H_L Q)^{-1}Q V_L P,
$
being generically non local, complex and energy dependent. If $V$ fulfills Hermiticity or parity symmetry, it follows that \cite{Ruschhaupt2017} $|T|^2=|\widetilde T|^2$ and $|R|^2=|\widetilde R|^2$ and from this we get $\cD = 0$. If $\cal V$ fulfills time reversal pseudohermiticity or PT symmetry, it follows that \cite{Ruschhaupt2017} $|T|^2=|\widetilde T|^2$ and from this we get $\cD = |\widetilde R|^2 -|R|^2$.
Finally, if $\cal V$ fulfills time reversal invariance or PT pseudohermiticity symmetry, it follows that $|R|^2=|\widetilde R|^2$ and from this we get $\cD = |\widetilde T|^2 -|T|^2$.

\begin{figure}
\begin{center}
(a) \includegraphics[width=0.43\linewidth]{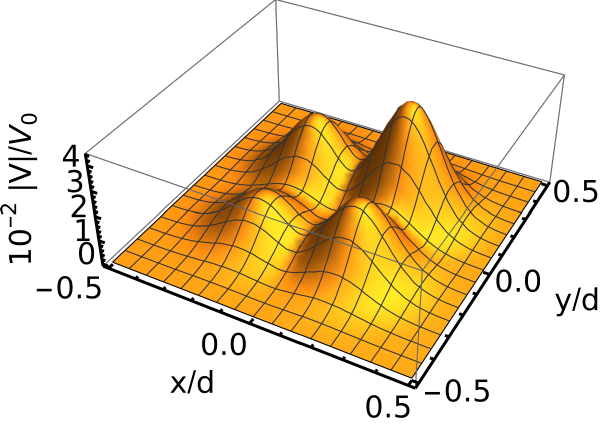}
(b) \includegraphics[width=0.43\linewidth]{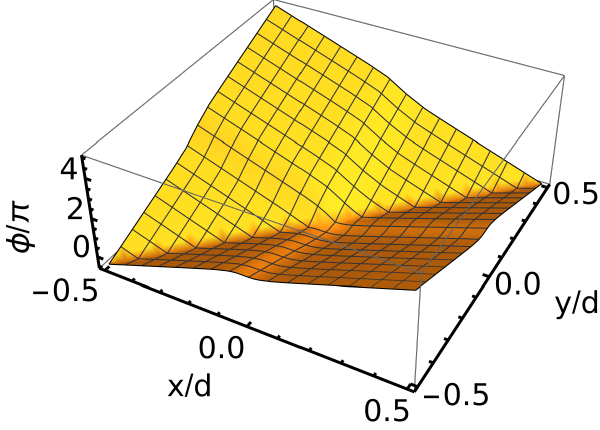}
\end{center}
\caption{Nonlocal potentials $V(x,y)$, see eq. \eqref{effpot}, for "half-demon" device ($|T|^2=\frac{1}{2}=|\widetilde R|^2$, $|\widetilde T|=|R|=0$ resulting in
$\cD=\frac{1}{2}$) for $v_0/v_d = 8$:
(a) absolute value, (b) argument. $b \tau=165.874$, $c \tau=103.876$, $x_0/d=0.16455$, $\Delta\tau=91.211$.\label{fig1}
}
\end{figure}

%
{\bf Boundary cases $\mathbf{\cD=\pm\frac{1}{2}}$.}
Let us now examine the upper bound case $\cD=\frac{1}{2}$ in more detail. It follows by taking also into account
eq. \eqref{a4} (left) that $0 \le |\widetilde T|^2 + |R|^2 = |T|^2 + |\widetilde R|^2 - 1 \le 0$. Therefore we
get that $|\widetilde T|=|R|=0$ and $|T|^2 + |\widetilde R|^2 = 1$ is required for $\cD=\frac{1}{2}$. This result also shows that $\cD=\frac{1}{2}$ is a tight upper bound. These are the devices which are closest to a coherent Maxwell demon. They include the interesting case
of a ``half-demon" with $|T|^2=\frac{1}{2}=|\widetilde R|^2$ (and $|\widetilde T|=|R|=0$).
Extending the method of \cite{Ruschhaupt2020}, we can implement such a ``half-demon'' or $\frac{1}{2}\cT/\frac{1}{2}\cR$ device using a two-level atom and a laser. By this, we achieve an effective non-local scattering potential for the ground-state where its kernel has the form
\beqa
\hspace*{-.3cm}V (x,y) 
= \frac{m}{4} \frac{e^{i|x-y|q}}{i q}
\Omega(x)\Omega(y)^*,
\label{effpot}
\eeqa
where  
$
q=\frac{\sqrt{2mE}}{\hbar}(1+\mu)^{1/2},\;\; 
{\rm Im}\,q\ge 0,
\label{qeq}
$ and 
$
\mu=\frac{2\Delta+i\gamma}{2E/\hbar}.
$
Here $E=\hbar^2 k^2/2m$ is the energy, and 
$\Omega(x)$ is the position-dependent, on-resonance Rabi frequency;
$\gamma$ is the inverse of the life time of the excited state and
$\Delta=\omega_{L}-\omega_{12}$
is the detuning (laser angular frequency minus the atomic transition 
angular frequency $\omega_{12}$).
We set now $\gamma=0$ and assume for the Rabi frequency the form
$\Omega (x) =  - i b g(x+x_0) + c g(x-x_0)$ (with no symmetry apart from the trivial symmetry) and with smooth, realizable Gaussians $g(x) =  \exp[-{x^2}/{w^2}]$.
We fix $2 d$ as an effective finite width 
of the potential area beyond which the potential is negligible and assumed to vanish.
We set $v_{d} = {\hbar}/({m d})$, $\tau={m d^2}/{\hbar}$, $V_0=\hbar^2/(md^3)$
and
we fix the width of the Gaussians to be $w= {\sqrt{2}}d/{10}$.
In the following, we set a target velocity $v_0=8 v_d$ to achieve the desired asymmetric scattering response, then the real parameters $b$, $c$, $x_0$ and  $\Delta$    
are numerically optimized. The resulting effective potential is shown in Fig. \ref{fig1}.

Other devices leading to $\cD=+\frac{1}{2}$ are the one-way T-filter ${\cal T/A}$ (an ``isolator'' in the language used for optical devices \cite{Mekawy2021}), and the one-way R-filter ${\cal A/R}$. Both devices have been designed using the $2$-level system crossing a laser beam in a quantum-optical setting mentioned above in \cite{Ruschhaupt2020}.

In a similar way, for the lower bound  $\cD=-\frac{1}{2}$, taking into account eq. \eqref{a4} (left) we get 
that 
$|T|=|\widetilde R|=0$ 
and $|\widetilde T|^2 + |R|^2 = 1$. 
This also shows that $\cD=-\frac{1}{2}$
is a tight lower bound.

\begin{figure}
(a)\includegraphics[width=0.35\linewidth]{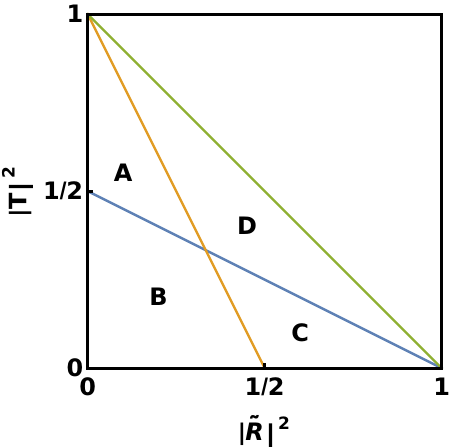}
(b)\includegraphics[width=0.5\linewidth]{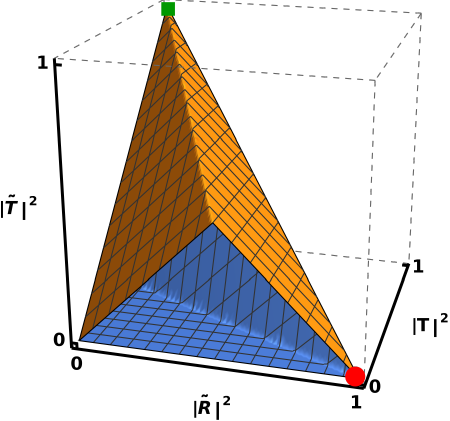}
\caption{
Set of allowed probabilities leading to $\cD=0$
(i.e., symmetrical scattering for right or left incidence (maximally away from demonic behavior)).
(a) different regions with different lower and upper bounds for $|\widetilde T|^2$ (see text);
(b) visualization of the lower bound (blue surface) and upper bound (orange surface) for
$|\widetilde T|^2$  (see text).\label{fig2}
}
\end{figure}

%
{\bf Case $\mathbf{\cD=0}$.}
Finally, let us also consider $\cD=0$, which corresponds to systems maximally away from any demon $\cD=\pm1$.
In this case we have together with eqs. \eqref{a4} that $0 \le |T|^2 + |\widetilde R|^2 = |\widetilde T|^2 + |R|^2 \le 1$, i.e., a symmetric response for left or right incidence.
In addition, the inequalities \eqref{a2} must be fulfilled. These inequalities can be then also rewritten as
$0 \le |\widetilde R|^2 \le 1 - |T|^2$ and $\max(0, 2 |T|^2 + |\widetilde R|^2 - 1) \le |\widetilde T|^2 \le \min (1-|\widetilde{R}|^2, |T|^2 + |\widetilde R|^2)$.
The resulting different regions with different upper and lower bounds for $|\widetilde T|^2$ can be seen in Fig. \ref{fig2}(a) [region A: $0 \le |\widetilde T|^2 \le 1-|\widetilde{R}|^2$, region B: $0 \le |\widetilde T|^2 \le |T|^2 + |\widetilde R|^2$, region C: $2 |T|^2 + |\widetilde R|^2 - 1 \le |\widetilde T|^2 \le |T|^2 + |\widetilde R|^2$, region D: $2 |T|^2 + |\widetilde R|^2 - 1 \le |\widetilde T|^2 \le 1-|\widetilde{R}|^2$], for a visualisation of these upper and lower bounds of $|\widetilde T|^2$ see also Fig. \ref{fig2}(b).
This class $\cD=0$ includes also the example of no potential ($|T|^2=1, |R|^2=0$ and $|\widetilde T|^2=1, |\widetilde R|^2=0$, green box in Fig. \ref{fig2}(b))
and the example of a full-reflecting potential ($|T|^2=0, |R|^2=1$ and $|\widetilde T|^2=0, |\widetilde R|^2=1$, red circle box in Fig. \ref{fig2}(b)).
Note that a device with $\cD=0$ does not need to be symmetric concerning parity. A counter-example would be a partial $\cal TR/A$ device 
($|T|^2=|R|^2=\frac{1}{2}$ and $|\widetilde T|^2=|\widetilde R|^2=0$) which would lead also to $\cD=0$.
Such a device has been designed using the $2$-level system crossing a laser beam in a quantum-optical setting mentioned above in \cite{Ruschhaupt2020}.

%
{\bf Demon-like devices by relaxing the requirements.}
In the previous section, we have proved that it is not possible to get a Maxwell demon (\ref{MDconditions}) for a given channel, irrespective of the
complexity of the large underlying system and its multichannel Hermitian Hamiltonian.  It is not even possible to get close to it.  
If we still want to construct demon-like devices, we have to relax the constraints imposed on the Maxwell demon so far.

{\it -Relaxing probability one:} One possibility is to relax the requirement that there is full transmission from the left and full reflection from the right.  In  the half-demon described above, there is a demonic response for half of the incidences, while for the other half the particles would end up in a different channel (i.e. they would be "absorbed").

{\it -Relaxing exit state:} Other possibility is to restrict the particle initially in a fixed state, e.g. the ground state, while the reflected or transmitted particle can be in an arbitrary state.
An `atom diode'' as described in refs. \cite{Ruschhaupt2004a,Ruschhaupt2006a,Ruschhaupt2006b} is essentially a Maxwell demon of this sort, in which ground state particles incident from the left are  transmitted to the right in an excited state, whereas the ground state particles  are reflected in the ground state when incident from the right. Strictly speaking these are  ${\cal{A/R}}$ devices where the ${\cal A}$ corresponds here to a transmitted excited state. Of course, this type of demon amounts to introducing some "asymmetry" in the defining constraints for incidence (in a ground state) and response (in any state). 
Extending the demon definition by allowing passage into a different channel and restricting the incident channels
(In optics this type of has been termed ``asymmetric transmission in reciprocal structures'' \cite{Sheikh2023})
may be in fact a rather generic mechanism to explain demonic-like behavior, in particular when the motional degree of freedom is the only relevant one and
the selection of a particular subspace for incidence occurs naturally.

{\it -Incoherent demons:} An atom diode \cite{Ruschhaupt2004a,Ruschhaupt2006a,Ruschhaupt2006b} as described above may be complemented by a mechanism to make the transmitted excited state decay incoherently to the ground state. The decay may be naturally occurring or forced by artificial quenching.  Note that incidence from the right in the excited state would lead to transmission but this type of incidence may be avoided by natural or forced decay. The resulting "demon", unlike the original definition in (1), is incoherent, described by an open system master equation which is outside the coherent assumptions of this paper as we are assuming a Schr\"odinger equation rather than a master equation as the starting point.

In this regard it is worth recalling the important contribution by Skordos and Zurek \cite{Skordos1992}.
In a numerical experiment they connected two gas chambers of equal volume through an asymmetric trapdoor that could only slide on rails in
one of the chambers. When the motion of the door was cooled, a pressure differential emerged between the two chambers, but not under normal conditions. In general, automated demons like this operate by capitalizing on dissipation.

%
{\bf Outlook.}
In this paper, we have assumed a linear Schr\"odinger equation of a large quantum system to find a no-go theorem to the implementation of asymmetric scattering devices as a Maxwell demon. We have also provided a lower bound for the quantified distance $\cE$ of any asymmetric scattering device to a Maxwell demon.

As stated above, non-linearities may change or break down the
symmetry selection rules. For example, non-linear effects may indeed be very relevant in optics leading to demonic amplitudes \cite{Peng2014,Riera2019, Mekawy2021}.
Therefore, this might be a direction of future research.

Moreover, we have assumed autonomous, time-independent Hamiltonians throughout, but the ensuing selection rules may
also be violated by spatio-temporal modulations that could lead in particular to different types of asymmetrical scattering. Much
progress has been achieved in optics in this regard  \cite{Sheikh2023}, but a similar research for particle scattering is yet to be
theoretically and technically developed, e.g. to find out the
spatio-temporal perturbations needed and the way to implement them. Some inspiring work in this line has been done in the related
field of asymmetric heat transport \cite{Riera2019}.

\begin{acknowledgments}
	A.R acknowledge that this publication has emanated from research
	funded by Taighde \'Eireann – Research Ireland
	under Grant Number 19/FFP/6951.
	J.G.M. thanks the Grant PID2021-126273NB-I00 funded by MCIN/AEI/ 10.13039/501100011033 and by “ERDF A way of making Europe”.
	J.G.M. acknowledges financial support from the Basque Government Grant No. IT1470-22.
\end{acknowledgments}

%
%


\begin{thebibliography}{10}

\bibitem{Maxwell1990}
{\it Maxwell's demon: Entropy, information, computing},
\newblock {P}rinceton University Press, Princeton, ed. by H. S. Leff and A. F.
  Rex (1990).

\bibitem{Maxwell2002}
{\it Maxwell demon 2 entropy, classical and quantum information, computing},
\newblock {CRC} Press, Boca Raton. ed. by H. S. Leff and A. F. Rex (2002).

\bibitem{Rex2017}
A.~Rex, {\em ``Maxwell’s demon—a historical review,''} Entropy {\bf 19},
  240 (2017).

\bibitem{Parrondo2021}
H.~Linke and J.~M.~R. Parrondo, {\em ``Tuning up maxwell’s demon,''} {
  Proceedings of the National Academy of Sciences} {\bf 118},
  e2108218118 (2021).

\bibitem{Koski2015}
J.~V. Koski, A.~Kutvonen, I.~M. Khaymovich, T.~Ala-Nissila, and J.~P. Pekola,
  {\em ``On-chip maxwell's demon as an information-powered refrigerator,''} {
  Phys. Rev. Lett.} {\bf 115}, 260602 (2015).

\bibitem{Leigh2007}
V.~Serreli, C.~F. Lee, E.~Kay, and L.~D. A., {\em ``A molecular information
  ratchet,''} {Nature} {\bf 445}, 523–527 (2007).

\bibitem{Mayusama2018}
Y.~Masuyama, K.~Funo, Y.~Murashita, A.~Noguchi, S.~Kono, Y.~Tabuchi,
  R.~Yamazaki, M.~Ueda, and Y.~Nakamura, {\em``Information-to-work conversion by
  maxwell’s demon in a superconducting circuit quantum electrodynamical
  system,''} {Nature Communications} {\bf 9}, 1291 (2018).

\bibitem{Ruschhaupt2004a}
A.~Ruschhaupt and J.~G. Muga, {\em ``Atom diode: A laser device for a unidirectional
  transmission of ground-state atoms,''} {Phys. Rev. A} {\bf 70}, 061604 (2004).

\bibitem{Ruschhaupt2006a}
A.~Ruschhaupt and J.~G. Muga, {\em ``Adiabatic interpretation of a two-level atom
  diode, a laser device for unidirectional transmission of ground-state
  atoms,''} {Phys. Rev. A} {\bf 73}, 013608 (2006).

\bibitem{Ruschhaupt2006b}
A.~Ruschhaupt, J.~G. Muga, and M.~G. Raizen, {\em ``Improvement by laser quenching
  of an ‘atom diode’: a one-way barrier for ultra-cold atoms,''} {J. Phys. B} {\bf 39}, L133 (2006).

\bibitem{Raizen2005}
M.~G. Raizen, A.~M. Dudarev, Q.~Niu, and N.~J. Fisch, {\em ``Compression of atomic
  phase space using an asymmetric one-way barrier,''} {Phys. Rev. Lett.} {\bf 94}, 053003 (2005).

\bibitem{Raizen2009}
M.~G. Raizen, {\em ``Comprehensive Control of Atomic Motion,''} Science {\bf 324}, 1403-1406 (2009).

\bibitem{Sheikh2023}
A.~S. Ansari, A.~K. Iyer, and B.~Gholipour, {\em ``Asymmetric transmission in
  nanophotonics,''} {Nanophotonics} {\bf 12}, 2639-2667 (2023).

\bibitem{Muga2004}
J.~Muga, J.~Palao, B.~Navarro, and I.~Egusquiza, {\em ``Complex absorbing
  potentials,''} {Physics Reports} {\bf 395}, 357--426 (2004).
\newblock Eq. (113) should read $\la x|V|x'\ra=\la-x|V|-x'\ra^*$.

\bibitem{Mosta2014}
A.~Mostafazadeh, {\em ``Transfer matrices as nonunitary $s$ matrices, multimode
  unidirectional invisibility, and perturbative inverse scattering,''} {Phys. Rev. A} {\bf 89}, 012709 (2014).

\bibitem{Ruschhaupt2017}
A.~Ruschhaupt, T.~Dowdall, M.~A. Sim{\'o}n, and J.~G. Muga, {\em ``Asymmetric
  scattering by non-hermitian potentials,''} {EPL (Europhysics Letters)}
  {\bf 120}, 20001 (2017).

\bibitem{Simon2018}
M.~A. Sim\'on, A.~Buend\'{\i}a, and J.~G. Muga, {\em ``Symmetries and invariants for
  non-hermitian Hamiltonians,''} {Mathematics} {\bf 6}, 111 (2018).

\bibitem{Simon2019}
M.~A. Sim\'on, A.~Buend\'{\i}a, A.~Kiely, A.~Mostafazadeh, and J.~G. Muga,
  {\em ``$s$-matrix pole symmetries for non-hermitian scattering Hamiltonians,''}
  {Phys. Rev. A} {\bf 99}, 052110 (2019).

\bibitem{Ruschhaupt2021}
A.~Ruschhaupt, M.~A. Simon, A.~Kiely, and J.~G. Muga, {\em ``The role of symmetry in
  non-hermitian scattering,''} {Journal of Physics: Conference Series} {\bf 2038}, 012020 (2021).

\bibitem{Lin2011}
Z.~Lin, H.~Ramezani, T.~Eichelkraut, T.~Kottos, H.~Cao, and D.~N.
  Christodoulides, {\em ``Unidirectional invisibility induced by
  $\mathcal{P}\mathcal{T}$-symmetric periodic structures,''} {Phys. Rev.
  Lett.} {\bf 106}, 213901 (2011).

\bibitem{Peng2014}
B.~Peng, S.~K. Özdemir, F.~Lei, F.~Monifi, G.~M., G.~L. Long, S.~Fan, F.~Nori,
  C.~Bender, and L.~Yang, {\em ``Parity–time-symmetric whispering-gallery
  microcavities,''} {Nature Phys.} {\bf 10}, 394 (2014).

\bibitem{Mosta2019}
A.~Mostafazadeh, {\em ``Nonlinear scattering and its transfer matrix formulation in
  one dimension,''} {The European Physical Journal Plus} {\bf 134}, 16 (2019).

\bibitem{Mekawy2021}
A.~Mekawy, D.~L. Sounas, and A.~Alù, {\em ``Free-space nonreciprocal transmission
  based on nonlinear coupled fano metasurfaces,''} {Photonics} {\bf 8}, 139 (2021).

\bibitem{Wigner1959}
E.~P. Wigner, {\em Group Theory and Its Application to the Quantum Mechanics of
  Atomic Spectra}.
\newblock New York, NY, USA: Academic Press (1959).

\bibitem{Feshbach1958}
H.~Feshbach, {\em ``Unified theory of nuclear reactions,''} {Annals of Physics} {\bf 5}, 357-390 (1958).

\bibitem{Feshbach1962}
H.~Feshbach, {\em ``A unified theory of nuclear reactions. ii,''} {Annals of
  Physics} {\bf 19}, 287-313 (1962).

\bibitem{Ruschhaupt2004b}
A.~Ruschhaupt, J.~A. Damborenea, B.~Navarro, J.~G. Muga, and G.~C. Hegerfeldt,
 {\em ``Exact and approximate complex potentials for modelling time observables,''}
  {EPL (Europhysics Letters)} {\bf 67}, 1 (2004).

\bibitem{Ruschhaupt2020}
A.~Ruschhaupt, A.~Kiely, M.~A. Sim\'on, and J.~G. Muga, {\em ``Quantum-optical
  implementation of non-hermitian potentials for asymmetric scattering,''} {Phys. Rev. A} {\bf 102}, 053705 (2020).

\bibitem{Skordos1992}
P.~A. Skordos and W.~H. Zurek, {\em ``Maxwell’s demon, rectifiers, and the second
  law: Computer simulation of Smoluchowski’s trapdoor,''} {American
  Journal of Physics} {\bf 60}, 876-882 (1992).

\bibitem{Riera2019}
A.~Riera-Campeny, M.~Mehboudi, M.~Pons, and A.~Sanpera, {\em ``Dynamically induced
  heat rectification in quantum systems,''} {Phys. Rev. E} {\bf 99},
  032126 (2019).

\end{thebibliography}
\end{document}